
\newif\ifPubSub \PubSubfalse
\newcount\PubSubMag \PubSubMag=1200
\def\PubSub{\PubSubtrue
            \magnification=\PubSubMag \hoffset=0pt \voffset=0pt
            \pretolerance=600 \tolerance=1200 \vbadness=1000 \hfuzz=3 true pt
            \baselineskip=1.75\baselineskip plus.2pt minus.1pt
            \parskip=2pt plus 6pt \gapskip=1.5\baselineskip
            \setbox\strutbox=\hbox{\vrule height .75\baselineskip
                                               depth  .25\baselineskip
                                               width 0pt}%
            \Page{5.75 true in}{8.9 true in}}

\newcount\FigNo \FigNo=0
\newbox\CapBox
\newbox\FigBox
\newtoks\0
\newif\ifImbed \Imbedfalse
\def\Fig {fig.~\the\FigNo}
\def\NFig {{\count255=\FigNo \advance\count255 by 1
            fig.\thinspace\the\count255}}

\def\StartFigure #1#2#3{\global\advance\FigNo by 1
                  \ifPubSub \global\setbox\FigBox=\vbox\bgroup
                            \unvbox\FigBox
                            \parindent=0pt \parskip0pt
                            \eject \line{\hfil}\bigskip\bigskip
                  \else \midinsert \removelastskip \vskip2\bigskipamount
                  \fi
                  \begingroup \hfuzz1in
                  \dimen0=\hsize \advance\dimen0 by-2\parindent \indent
                  \vbox\bgroup\hsize=\dimen0 \parindent=0pt \parskip=0pt
                       \vrule height1pt depth0pt width0pt
                       \ifdim\dimen0<#1bp \dimen1=#1bp
                                          \advance\dimen1 by -\dimen0
                                          \divide\dimen1 by 2 \hskip-\dimen1
                       \fi
                       \hfil
                       \vbox to #2 bp\bgroup\hsize #1 bp
                            \vss\noindent\strut\special{"#3}%
                            \skip0=\parskip \advance\skip0 by\dp\strutbox
                            \vskip-\skip0 }%
\def\caption #1{\strut\egroup
                \ifPubSub \global\setbox\CapBox=\vbox{\unvbox\CapBox
                                        \parindent0pt \medskip
                                         {\bf Figure \the\FigNo:}
                                          {\tenrm\strut #1\strut}}%
                \else \bigskip {\FootCapFace \hfuzz=1pt \baselineskip=3ex
                                \noindent{\bf Figure~\the\FigNo:} #1\par}%
                \fi}%
\def\label (#1,#2)#3{{\offinterlineskip \parindent=0pt \parskip=0pt
                     \vskip-\parskip 
                     \vbox to 0pt{\vss
                          \moveright #1bp\hbox to 0pt{\raise #2bp
                                         \hbox{#3}\hss}\hskip-#1bp\relax}}}%
\def\EndFigure {\egroup \endgroup 
                \ifPubSub \vfil
                          \centerline{\bf Figure \the\FigNo}
                          \egroup 
                \else \bigskip \endinsert 
                \fi \Imbedfalse}
\def\ListCaptions {\vfil\eject \message{!Figure captions:!}%
                   \Sectionvar{Figure Captions}\par
                   \unvbox\CapBox}
\def\ShowFigures {\ifPubSub\ifnum\FigNo>0 \ListCaptions \vfil\eject
                                 \message{!Figures:!}%
                                 \nopagenumbers \unvbox\FigBox \eject
                           \fi
                  \fi}
\def\FootCapFace{} 


\parskip=0pt plus 3pt
\def\Page #1#2{{\dimen0=\hsize \advance\dimen0 by-#1 \divide\dimen0 by 2
               \global\advance\hoffset by \dimen0
               \dimen0=\vsize \advance\dimen0 by-#2 \divide\dimen0 by 4
               \ifdim\dimen0<0pt \multiply\dimen0 by 3 \fi
               \global\advance\voffset by \dimen0
               \global\hsize=#1 \global\vsize=#2\relax}
               \ifdim\hsize<5.5in \tolerance=300 \pretolerance=300 \fi}
\def\EndPaper{\par\dosupereject \ifnum\RefNo>1 \ShowReferences \fi
              \ShowFigures
              \par\vfill\supereject
              \message{!!That's it.}\end}
\headline{\ifnum\pageno=-1 \hfil \Smallrm \Time, \Date \else \hfil \fi}
\def\VersionInfo #1{\headline{\ifnum\pageno=-1 \hfil \Smallrm #1
                              \else \hfil
                              \fi}}


\newcount\RefNo \RefNo=1
\newbox\RefBox
\def\Jou #1{{\it #1}}
\def\Vol #1{{\bf #1}}
\def\AddRef #1{\setbox\RefBox=\vbox{\unvbox\RefBox
                       \parindent1.75em
                       \pretolerance=1000 \hbadness=1000
                       \vskip0pt plus1pt
                       \item{\the\RefNo.}
                       \sfcode`\.=1000 \strut#1\strut}%
               \global\advance\RefNo by1 }
\def\Ref  #1{\Hunskip~[{\the\RefNo}]\AddRef{#1}}
\def\Refc #1{\Hunskip~[{\the\RefNo,$\,$}\AddRef{#1}}
\def\Refm #1{\Hunskip{\the\RefNo,$\,$}\AddRef{#1}}
\def\Refe #1{\Hunskip{\the\RefNo}]\AddRef{#1}}
\def\Refl #1{\Hunskip~[{\the\RefNo}--\nobreak\AddRef{#1}}
\def\Refn #1{\Hunskip\AddRef{#1}}
\def\ShowReferences {\message{!References:!}%
                     \Sectionvar{References}\par
                     \vskip-5\parskip
                     \unvbox\RefBox}
\def\StoreRef #1{\Hunskip\edef#1{\the\RefNo}}


\newif\ifRomanNum
\newcount \Sno \Sno=0
\def\Interskip #1#2#3{{\removelastskip \dimen0=#1
                       \advance\dimen0 by2\baselineskip
                       \vskip0pt plus\dimen0 \penalty-300
                       \vskip0pt plus-\dimen0
                       \advance\dimen0 by -2\baselineskip
                       \vskip\dimen0 plus#2 minus#3}}
\def\Section #1\par{\Interskip{24pt}{6pt}{2pt}%
                    \global\advance\Sno by1
                    \setbox0=\hbox{\STitlefont
                                    \ifRomanNum
                                       \global\SSno=64 \uppercase
                                       \expandafter{\romannumeral
                                                    \the\Sno.\ \ }%
                                     \else
                                        \global\SSno=96 \the\Sno.\ \
                                     \fi}
                    \leftline{\vtop{\copy0 }%
                              \vtop{\advance\hsize by -\wd0
                                    \raggedright
                                    \pretolerance10000 \hbadness10000
                                    \noindent \STitlefont
                                    \GetParenDim{\dimen0}{\dimen1}%
                                    \advance\dimen0 by\dimen1
                                    \baselineskip=\dimen0 #1}}
                    \hrule height0pt depth0pt
                    \dimen0=\baselineskip \advance\dimen0 by -\parskip
                    \nobreak\vskip\dimen0 plus 3pt minus3pt \noindent}%
\def\Sectionvar #1\par{\Interskip{24pt}{6pt}{2pt}%
                    \leftline{\vbox{\noindent
                                    \STitlefont #1}}
                    \hrule height0pt depth0pt
                    \dimen0=\baselineskip \advance\dimen0 by -\parskip
                    \nobreak\vskip\dimen0 plus 3pt minus3pt \noindent}%
\newcount\SSno \SSno=0
\def\SubSection #1\par{\Interskip{15pt}{3pt}{1pt}%
                    \global\advance\SSno by1
                    \setbox0=\hbox{\SSTitlefont \char\SSno$\,$]$\,$\ }
                    \leftline{\vtop{\copy0 }%
                              \vtop{\advance\hsize by -\wd0
                                    \raggedright
                                    \pretolerance10000 \hbadness10000
                                    \noindent \SSTitlefont
                                    \GetParenDim{\dimen0}{\dimen1}%
                                    \advance\dimen0 by\dimen1
                                    \baselineskip=\dimen0 #1}}
                    \hrule height0pt depth0pt
                    \dimen0=.8\baselineskip \advance\dimen0 by -2\parskip
                    \nobreak\vskip\dimen0 plus1pt minus1pt \noindent}%
\def\SubSectionvar #1\par{\Interskip{15pt}{3pt}{1pt}%
                    \leftline{\vbox{\noindent
                                    \SSTitlefont #1}}
                    \hrule height0pt depth0pt
                    \dimen0=.8\baselineskip \advance\dimen0 by -2\parskip
                    \nobreak\vskip\dimen0 plus 1pt minus1pt \noindent}%
\def\(#1){~\ifRomanNum {\Medrm\uppercase\expandafter{\romannumeral #1}}%
           \else {#1}%
           \fi}


\newcount\EqNo \EqNo=0
\def\NumbEq {\global\advance\EqNo by 1
             \eqno(\the\EqNo)}
\def\PrevEq {(\the\EqNo)}
\def\PrevEqs #1{{\count255=\EqNo \advance\count255 by-#1\relax
                 (\the\count255)}}
\def\NameEq #1{\xdef#1{(\the\EqNo)}\ignorespaces}

\def\AppndEq #1{\EqNo=0
  \def\NumbEq {\global\advance\EqNo by 1
             \eqno(#1\the\EqNo)}
  \def\numbeq {\global\advance\EqNo by 1
             (#1\the\EqNo)}
  \def\PrevEq {(#1\the\EqNo)}
  \def\PrevEqs ##1{{\count255=\EqNo \advance\count255 by-##1\relax
                 (#1\the\count255)}}
  \def\NameEq ##1{\xdef##1{(#1\the\EqNo)}\ignorespaces}}


\newcount\ThmNo \ThmNo=0
\newcount\LemNo \LemNo=0
\def\Theorem #1\par{\removelastskip\bigbreak
    \advance\ThmNo by 1
    \noindent{\bf Theorem \the\ThmNo:} {\sl #1\bigskip}}
\def\Lemma #1\par{\removelastskip\bigbreak
    \advance\LemNo by 1
    \noindent{\bf Lemma \the\LemNo:} {\sl #1\bigskip}}

\def\PrevThm {Theorem~\the\ThmNo}
\def\Thm {{\advance\ThmNo by 1 \PrevThm}}
\def\PrevLm {lemma~\the\LemNo}
\def\Declare #1#2\par{\removelastskip\bigbreak
    \noindent{\bf #1:} {\sl #2\bigskip}}


\newlinechar=`\!
\def\\{\ifhmode\hfil\break\fi}
\let\thsp=\,
\def\,{\ifmmode\thsp\else,\thinspace\fi}
\def\Hunskip {\ifhmode\unskip\fi}

\def\Date {\ifcase\month\or January\or February\or March\or April\or
 May\or June\or July\or August\or September\or October\or November\or
 December\fi\ \number\day, \number\year}

\newcount\mins  \newcount\hours
\def\Time{\hours=\time \mins=\time
     \divide\hours by60 \multiply\hours by60 \advance\mins by-\hours
     \divide\hours by60         
     \ifnum\hours=12 12:\ifnum\mins<10 0\fi\number\mins~P.M.\else
       \ifnum\hours>12 \advance\hours by-12 
         \number\hours:\ifnum\mins<10 0\fi\number\mins~P.M.\else
            \ifnum\hours=0 \hours=12 \fi
         \number\hours:\ifnum\mins<10 0\fi\number\mins~A.M.\fi
     \fi }

\def\HollowBox #1#2#3{{\dimen0=#1
       \advance\dimen0 by -#3 \dimen1=\dimen0 \advance\dimen1 by -#3
        \vrule height #1 depth #3 width #3
        \hskip -#3
        \vrule height 0pt depth #3 width #2
        \llap{\vrule height #1 depth -\dimen0 width #2}%
       \hskip -#3
       \vrule height #1 depth #3 width #3}}
\newbox\Tombstone
\setbox\Tombstone=\vbox{\hbox{\HollowBox{8pt}{5pt}{.8pt}}}

\def\GetParenDim #1#2{\setbox1=\hbox{(}%
                      #1=\ht1 \advance#1 by 1pt
                      #2=\dp1 \advance#2 by 1pt}

\def\Bull #1#2\par{\removelastskip\smallbreak\textindent{$\bullet$}
                    {\it #1}#2\smallskip}


\def\Title #1\par{\message{!!#1!}%
                  \nopagenumbers \pageno=-1
                  {\leftskip=0pt plus 1fil \rightskip=0pt plus 1fil
                   \parfillskip=0pt \Titlefont
                   \ifdim\baselineskip<2.5ex \baselineskip=2.5ex \fi
                   \noindent #1\par}\bigskip\bigskip\bigskip}
\def\Author #1\par{{\count1=0   
                    \count2=0   
                    \dimen0=0pt
                    \Position{\ignorespaces\indent
                              #1\hskip-\dimen0 }\bigskip}}
\def\Address #1\par{{\leftskip=0pt plus 1fil \rightskip=0pt plus 1fil
                     \parindent=0pt \parfillskip=0pt 
                     \ifdim\baselineskip>3ex \baselineskip=2.7ex \fi
                     \sl #1\bigskip}}
\def\modfnote #1#2{{\parindent=1.1em \leftskip=0pt \rightskip=0pt
                    \GetParenDim{\dimen1}{\dimen2}%
                    \setbox0=\hbox{\vrule height\dimen1 depth\dimen2
                                          width 0pt}%
                    \advance\dimen1 by \dimen2 \baselineskip=\dimen1
                    \vfootnote{#1}{\hangindent=\parindent \hangafter=1
                                 \unhcopy0 #2\unhbox0
                                 \vskip-\gapskip }}}%
\def\PAddress #1{\ifcase\count1 \let\symbol=\dag
                 \or \let\symbol=\ddag
                 \or \let\symbol=\P
                 \or \let\symbol=\S
                 \else \advance\count1 by -3
                       \def\symbol{\dag_{\the\count1 }}%
                 \fi
                 \advance\count1 by 1
                 \setbox0=\hbox{$^{\symbol}$}\advance\dimen0 by .5\wd0
                 \Hunskip \box0
                 \modfnote{$\symbol$}{{\sl Permanent address\/}: #1}}%
\def\Email #1{\ifcase\count2 \let\symbol=\ast
                 \or \let\symbol=\star
                 \or \let\symbol=\bullet
                 \or \let\symbol=\diamond
                 \or \let\symbol=\circ
                 \else \advance\count2 by -4
                       \def\symbol{\ast_{\the\count2 }}%
                 \fi
                 \advance\count2 by 1
                 \setbox0=\hbox{$^{\textstyle\symbol}$}\advance\dimen0 by .5\wd0
                 \Hunskip \box0
                 \modfnote{$\symbol$}{{\sl Electronic mail\/}: #1}}%
\def\Abstract{\vfil \message{!Abstract:!}%
              \dimen0=\hsize \advance \dimen0 by-2\parindent
              \setbox0=\vbox\bgroup\hsize=\dimen0
              \Position{{\STitlefont Abstract}}
              \smallskip
              \ifdim\baselineskip>4.5ex \baselineskip=4.5ex \fi
              \noindent\ignorespaces}%
\def\EndAbstract{\par\egroup\noindent\hfil\shade{\copy0 }\par}
\def\shade #1{{\setbox1=\hbox{#1}\dimen0=\wd1
               \dimen1=\ht1 \advance\dimen1 by \dp1
               \advance\dimen0 by30pt
               \count255=\dimen0 \divide\count255 by 65536
               \advance\dimen1 by 10pt
               \count1=\dimen1 \divide\count1 by 65536
               \gray{\the\count255 }{\the\count1 }\hskip-\dimen0 \hskip14pt #1}}
\def\gray #1#2{\hbox to #1bp{
                                 \hss}}
\def\pacs #1{\vfil\leftline{PACS numbers: #1}\eject}
\def\StartPaper{\pageno=1 \ifPubSub \footline{\tenrm\hss
                                              --\ \folio\ -- \hss}%
                          \else \footline{\hss\vtop to 0pt{\hsize=.15\hsize
                                     \vglue6pt \hrule \medskip
                                     \centerline{\tenrm\folio}\vss}\hss}%
                          \fi}

\newskip\gapskip \gapskip=\baselineskip


\RomanNumtrue
\let\Position=\centerline 

\def\Titlefont{\tenbf}   
\def\STitlefont{\tenbf}  
\def\SSTitlefont{\tensl} 
\def\Smallrm{\sevenrm}   
\def\Medrm{\tenrm}       

\newcount\EZReadMag \EZReadMag=1200
\def\EZRead{\magnification=\EZReadMag \hoffset=0pt \voffset=0pt
            \pretolerance=1000 \tolerance=2000
            \vbadness=1000 \hbadness=1500 \hfuzz=4 true pt
            \baselineskip=1.1\baselineskip plus.1pt minus0pt
            \parskip=0pt plus 1pt \gapskip=1\baselineskip
            \setbox\strutbox=\hbox{\vrule height .75\baselineskip
                                               depth  .25\baselineskip
                                               width 0pt}%
            \Page{5.5 true in}{8.5 true in}}
\def\VEZRead{\magnification=\PubSubMag \hoffset=0pt \voffset=0pt
            \pretolerance=1000 \tolerance=2000
            \vbadness=1000 \hbadness=1000 \hfuzz=4 true pt
            \baselineskip=1.1\baselineskip plus.1pt minus0pt
            \parskip=0pt plus 1pt \gapskip=\baselineskip
            \setbox\strutbox=\hbox{\vrule height .75\baselineskip
                                               depth  .25\baselineskip
                                               width 0pt}%
            \Page{5 true in}{8 true in}}

\def\stacksymbols #1#2#3#4{\def\theguybelow{#2}
    \def\verticalposition{\lower#3pt}
    \def\spacingwithinsymbol{\baselineskip0pt\lineskip#4pt}
    \mathrel{\mathpalette\intermediary#1}}
\def\intermediary#1#2{\verticalposition\vbox{\spacingwithinsymbol
      \everycr={}\tabskip0pt 
      \halign{$\mathsurround0pt#1\hfil##\hfil$\crcr#2\crcr
               \theguybelow\crcr}}}

\def\lapproxeq{\stacksymbols{<}{\sim}{2.5}{.2}}

\def\grad{\nabla}


\message{!<*><*><*><*><*><*><*><*><*><*><*><*><*><*><*><*><*><*><*><*><*><*><*>}
\message{!PROCESSING NOTE:!
         If you print the dvi file via `dvips' the figures will!
         automatically emerge. If you use anything else, you will!
         get the text of the paper, but no figures.!
         Send e-mail to borde@cosmos2.phy.tufts.edu if you have any!
         problems.!!}
\message{<*><*><*><*><*><*><*><*><*><*><*><*><*><*><*><*><*><*><*><*><*><*><*>!}



\VEZRead
\VersionInfo{February 10, 1997}

\Title Violation of the Weak Energy Condition in~Inflating~Spacetimes

\Author Arvind Borde
\PAddress{Department of Mathematics, Southampton College, NY 11968.}
\Email{borde@cosmos2.phy.tufts.edu}
and Alexander Vilenkin
\Email{vilenkin@cosmos2.phy.tufts.edu}

\Address Institute of Cosmology, Department of Physics and Astronomy\\
         Tufts University, Medford, MA 02155, USA.

\Abstract
We argue that many future-eternal inflating spacetimes are likely to
violate the weak
energy condition.  It is possible that such spacetimes may not enforce
any of the known averaged conditions either.  If this is indeed the case,
it may open the door to constructing non-singular, past-eternal inflating 
cosmologies.  Simple non-singular models are, however, unsatisfactory,
and it is not clear if satisfactory models can be built that
solve the problem of the initial singularity
\EndAbstract


\StartPaper

\Section Introduction

Inflationary cosmological models 
\Ref{For reviews of inflation see, for example,\\
S.K. Blau and A.H. Guth, in \Jou{300 Years of Gravitation},
edited by S.W. Hawking and W. Israel (Cambridge University Press,
Cambridge, England, 1987);\\
A.D. Linde, \Jou{Particle Physics and Inflationary Cosmology\/} 
(Harwood Academic, Chur Switzerland, 1990);\\
E.W. Kolb and M.S. Turner, \Jou{The Early Universe\/} 
(Addison-Wesley, New York, 1990).}
are generically future-eternal
\StoreRef{\VilenkinEt}
\Refl{A. Vilenkin, \Jou{Phys. Rev.~D}, \Vol{27}, 2848 (1983).}
\StoreRef{\LindeOne}
\Refn{A.D. Linde, \Jou{Phys. Lett. B\/} \Vol{175}, 395 (1986).}
\StoreRef{\AryaVi}
\Refn{M. Aryal and A. Vilenkin, \Jou{Phys. Lett. B\/} \Vol{199}, 351 (1987).}
\StoreRef{\LindeTwo}
\Refn{A.S. Goncharov, A.D. Linde and V.F. Mukhanov,
        \Jou{Int. J. Mod. Phys. A\/} \Vol{2}, 561 (1987).}
\Refn{K. Nakao, Y. Nambu and M. Sasaki, \Jou{Prog. Theor. Phys.\/}
        \Vol{80}, 1041 (1988).}
\StoreRef{\LindeThree}
\Refe{A. Linde, D. Linde and A. Mezhlumian, \Jou{Phys. Rev.~D}, \Vol{49},
        1783 (1994).}.
In such models, the Universe consists of a number of
post-inflationary, thermalized regions 
embedded in an always-inflating background.  The thermalized regions
grow in time, but the inflating background in which they are embedded
grows even faster, and the thermalized regions do not, in general, merge.
As a result, there never arrives an instant of time after which the
Universe is completely thermalized.  This scenario is schematically
illustrated in \NFig.

\StartFigure{320}{100}
    {gsave
     .5 setlinewidth
     [1 1] 0 setdash
     newpath 33 85 moveto 48 60 58 60 73 85 curveto
             gsave .9 setgray fill grestore stroke
     newpath 93 85 moveto 148 20 158 20 213 85 curveto
             gsave .9 setgray fill grestore stroke
     newpath 223 85 moveto 235 67 241 67 253 85 curveto
             gsave .9 setgray fill grestore stroke
     newpath 256 85 moveto 281 50 291 50 316 85 curveto
             gsave .9 setgray fill grestore stroke
     1.5 setlinewidth
     newpath 30 15 moveto 320 15 lineto stroke
     newpath 30 85 moveto 320 85 lineto stroke
     grestore
     }
\label(0,40){$\matrix{\Bigg\uparrow\cr\hbox{time}\cr}$}
\label(0,92){\hbox to 370 bp{\hss future infinity\hss}}
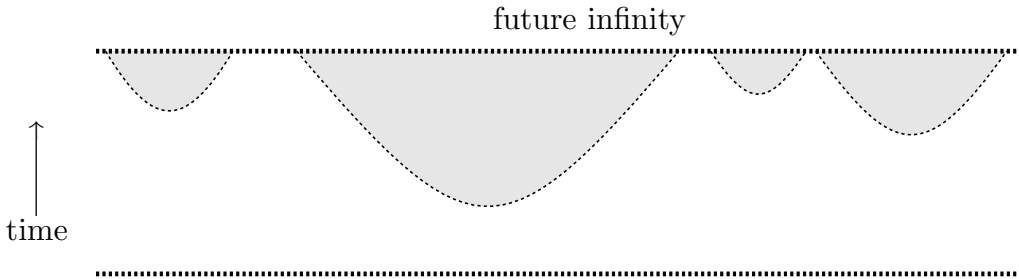
\caption{A schematic representation of an inflating Universe.
         The shaded regions are thermalized regions, where inflation
         has ended.  We live in such a region (i.e., the entire
         observed Universe lies within a single thermalized region).}
\EndFigure

Quantum fluctuations of the inflaton field $\phi$ play an essential role 
in many models of eternal inflation.  
In such models there is a parameter~$H$ 
(the Hubble parameter, also referred to as the expansion rate), such that  
the fluctuations of $\phi$ can be pictured as a ``random walk,'' 
or ``diffusion,'' in
which $\phi$ varies by approximately $\pm H/2\pi$ on the scale
$H^{-1}$  (the ``horizon scale'')
per time~$H^{-1}$ (the ``Hubble time'').  
The fluctuations are superimposed on the 
classical evolution of~$\phi$ determined by its potential~$V(\phi)$.
Although there is an overall tendency for~$\phi$ to roll down the potential,
it will be pushed up occasionally by quantum fluctuations.
It is this effect that is responsible for the eternal nature of inflation
\Ref{The quantum fluctuations make $\phi$ go up and down the potential.
This means that different parts of the
Universe thermalize at different times. Further,
at any given time, there
is a non-zero probability for some regions of the
Universe to still be inflating.}.

These quantum fluctuations of $\phi$ induce fluctuations of the spacetime 
geometry, and we expect that the expansion rate will also fluctuate
from one horizon-size region to another. The quantum nature of the 
fluctuations becomes unimportant when the expansion of the Universe
stretches their wavelength well beyond the horizon.
Hence, one can meaningfully define classical spacetime histories
for the scalar field $\phi^{\rm (av)}(x)$ and the metric 
$g^{\rm (av)}_{\mu\nu}(x)$  averaged (``smeared'') over a scale
$\ell>H^{-1}$
\Ref{In fact, the ``horizon size'' itself will be subject to fluctuations,
but is a meaningful concept when averaged over this large scale.}.

In the rest of this paper the 
spacetime geometry 
and the field $\phi$ will be understood in the averaged sense defined above,
and we shall drop the superscript ``(av)''.
In the inflating part of the Universe both the averaged field $\phi$
and the expansion rate $H$ are expected to be slowly varying functions,
i.e., $(\partial_{\mu}H)^2 \ll H^4$.

The future-eternal nature of inflation suggests that we consider
the possibility that inflating
spacetimes can also be extended to the infinite past, resulting
in a ``steady-state'' non-singular cosmological model.  
This possibility was discussed in the early days of inflation
\StoreRef{\EarlyU}
\Ref{See, for example, some of the discussions in \Jou{The Very Early Universe},
edited by G.W.~Gibbons and S.W.~Hawking, Cambridge University Press (1983).}
but it was soon realized by Linde
\Ref{A.D. Linde, in ref.~[\EarlyU.].}
and by others~[\VilenkinEt\,
\Refe{P.J. Steinhardt, in ref.~[\EarlyU.].}
that the idea could not be implemented in the simplest model in which
the inflating Universe is described by an exact de~Sitter space. 
It was then proved by one of us
\StoreRef{\Vilenkin} 
\Ref{A. Vilenkin, \Jou{Phys. Rev.~D}, \Vol{46}, 2355 (1992).}
that a generic 2-dimensional spacetime
that was eternally inflating to the future could not be geodesically
complete to the past
\Ref{A spacetime is {\it past-geodesically complete\/} if all timelike
and null geodesics can be extended in the past direction to 
infinite values of their affine parameters.}.      
This paper also gave a plausibility argument
that suggested that the 2-dimensional result would continue to 
hold in four spacetime dimensions. 

A rigorous four-dimensional proof
was subsequently provided by us
\StoreRef{\BVOne}
\Refc{A. Borde and A. Vilenkin, \Jou{Phys. Rev. Lett.}, \Vol{72}, 3305 
      (1994).}
\StoreRef{\BVTwo}
\Refe{A. Borde and A. Vilenkin, in \Jou{Relativistic Astrophysics: 
      The Proceedings 
      of the Eighth Yukawa Symposium}, edited by M.\ Sasaki, 
      Universal Academy Press, Japan (1995).}, 
in a theorem that showed that under some natural
assumptions about the spacetime geometry, a future-eternal
inflationary model cannot be globally extended into the
infinite past; 
i.e., it is not geodesically complete in the past direction.
The assumptions that lead to geodesic incompleteness
in this result are the following:

\bigskip
\item{A]} The Universe is causally simple
\Ref{This is the requirement that spacetime have a simple causal structure.
In particular, it excludes complicated topological interconnections
between different regions of spacetime. See~[\BVOne\,\BVTwo] for a  
precise discussion and a diagram.}.
(A theorem with this condition replaced by a condition
called the ``limited influence condition'' was subsequently obtained
\StoreRef{\BVThree}
\Refc{A. Borde and A. Vilenkin, \Jou{Int. J. Mod. Phys.}, to appear.}
\Refe{A. Borde, preprint (1996).}.)
\item{B]} The Universe is open.
(An extension to certain closed Universes was subsequently obtained
\StoreRef{\Borde}
\Ref{A. Borde, \Jou{Phys. Rev. D.}, \Vol{50}, 3392 (1994).}.)
\item{C]} The Universe
obeys the ``finite past-volume difference condition''
\Ref{This requires that there exist certain pairs of points
such that the spacetime volume of the difference of their
pasts is finite~-- a condition necessary for inflation
to persist in the future time direction. See~[\BVOne\,\BVTwo] for
details.}. 
\item{D]} The Universe obeys the null convergence condition. 

\bigskip
\noindent
The main purpose of the present paper is to
re-examine the validity of this last condition.

We use conventions in which Einstein's equation is
$$
R_{\mu\nu} - {\textstyle{1\over 2}} R g_{\mu\nu} = 
8\pi GT_{\mu\nu}.\NumbEq
$$
\NameEq{\Einstein}
Under these conventions, the
null convergence condition requires that the Ricci tensor,
$R_{\mu\nu}$, satisfy 
$$
R_{\mu\nu}N^\mu N^\nu \geq 0\NumbEq
$$ 
\NameEq{\NullConv}
for all null vectors~$N^\mu$.
This condition is closely related to the
weak energy condition, which requires that the energy-momentum tensor, 
$T_{\mu\nu}$, satisfy 
$$
T_{\mu\nu}V^\mu V^\nu \geq 0\NumbEq
$$
\NameEq{\WeakEn} 
for all timelike vectors~$V^\mu$.
An observer whose worldline has tangent $V^\mu$
at a point will measure an energy density of $T_{\mu\nu}V^\mu V^\nu$ at
that point.  Thus, the weak energy condition means physically that the 
matter energy density
is non-negative when measured by any observer.

In models that obey Einstein's equation~\Einstein\ 
a violation of the null convergence condition~\NullConv\ implies a
violation of  the weak energy condition~\WeakEn.
To see this, suppose that there is a (say, future-directed) null
vector, $N^\mu$, such that $R_{\mu\nu}N^\mu N^\nu=-\delta<0$. 
Einstein's equation~\Einstein\ implies that
$T_{\mu\nu}N^\mu N^\nu=-(8\pi G)^{-1}\delta<0$.
Then the timelike
vector given by $V^\mu =N^\mu+\epsilon T^\mu $, 
where $T^\mu $ is a unit, future-directed
timelike vector, will obey
$T_{\mu\nu}V^\mu V^\nu <0$ for sufficiently small values
of $\epsilon$.
      
Thus, the null convergence condition 
appears to be a very reasonable
requirement on the spacetime geometry.  For a perfect-fluid
spacetime with energy density~$\rho$ and pressure~$p$    
the weak energy condition (and, therefore, the null convergence
condition) holds if $\rho\geq 0$ and $\rho+p\geq 0$.
This is satisfied by all known forms of matter. 
An inflating Universe is characterized by a nearly vacuum
equation of state, $p\approx -\rho$, and, when the exact equality
holds, the null convergence condition~\NullConv\
is satisfied~-- but only marginally.  This is less unstable than it 
seems, because all classical deviations from the
vacuum equation of state appear to work in the direction of making
$\rho+p$ positive rather than negative.
For example, the energy-momentum tensor of the inflaton field
$\phi$ is
$$
T_{\mu\nu}=\partial_{\mu}\phi\partial_{\nu}\phi - g_{\mu\nu}
\left[{\textstyle{1\over 2}}(\partial_{\sigma}\phi)^2 - V(\phi)\right],
\NumbEq
$$
and we can write
$$
R_{\mu\nu}N^{\mu}N^{\nu} = 8 \pi G T_{\mu\nu}N^\mu N^\nu =
8 \pi G (N^{\mu}\partial_{\mu} \phi)^2
 \geq 0.
\NumbEq
$$
\NameEq{\NullInf}
Moreover, the addition of any ordinary matter with $p>0$ further tips the balance in
the direction of a positive sign for $R_{\mu\nu}N^{\mu}N^{\nu}$.   

Equation~\NullInf\ shows us
that the null convergence condition is satisfied in
inflationary models as long as their dynamics is accurately described by
Einstein's classical equation with a scalar field source.  The situation
is not so clear in the ``diffusion'' regions of spacetime where the dynamics
is dominated by quantum fluctuations of~$\phi$. The energy-momentum tensor
in such regions can be written as
$$
T_{\mu\nu} = T_{\mu\nu}[\phi] + T_{\mu\nu}^{\rm (fluct)},\NumbEq
$$
where $T_{\mu\nu}[\phi]$ is constructed from the smeared-over-an-horizon
field $\phi(x)$ and $T_{\mu\nu}^{\rm (fluct)}$ is the contribution
of short-wavelength modes of $\phi$ (with wavelengths $\lambda \lapproxeq H^{-1}$).
Accounting for this contribution in
a systematic way remains an interesting unsolved problem.  
For our purposes it will be sufficient to use the estimate
$$
T_{\mu\nu}^{\rm (fluct)}\sim H^4.\NumbEq
$$  
\NameEq{\Est}
(This estimate is easily understood
if we recall that $\phi$ fluctuates by $\delta\phi \sim H$
on time and length scales $\delta t \sim \delta \ell \sim H^{-1}$
and that $T_{\mu\nu}$ includes terms quadratic
in gradients of $\phi$.)  In the diffusion region, the smeared field gradients
are small (i.e.,  $\left|\partial_{\mu}\phi\right| \lapproxeq H^2$), and 
$T_{\mu\nu}N^\mu N^\nu$ will now contain both a manifestly non-negative term, as in
eqn.~\NullInf, as well as  a non-negligible correction from 
$T_{\mu\nu}^{\rm (fluct)}$.  It is no longer obvious, in this case, whether
the null convergence condition \NullConv\ is satisfied. 
In the next section we argue that the condition 
is indeed violated in the ``diffusion'' regions of inflationary spacetimes. 
This violation may open the door to escaping the conclusion of our previous 
theorems~[\BVOne\,\BVThree], and towards constructing past-eternal, 
non-singular cosmologies.
Violations of the weak energy condition may also allow us to avoid
the conclusions of Farhi and Guth
\Ref{E. Farhi and A.H. Guth, \Jou{Phys. Lett. B}, \Vol{183}, 149 
     (1987).} 
whose results appeared to forbid the creation of an inflating 
Universe in a laboratory. (Other ways
around the results of Farhi and Guth have previously appeared in the literature
\Refl{E. Farhi, A.H. Guth and J. Guven, \Jou{Nucl. Phys.}, \Vol{B339}, 417 
     (1990).}
\Refn{L. Fischler, D. Morgan and J. Polchinski, \Jou{Phys. Rev. D}, \Vol{41},
     2638 (1990).}
\Refe{A. Linde, \Jou{Nucl. Phys}, \Vol{B372}, 421 (1992).}.)     

The rest of this paper is organized as follows: In section~II we discuss
how the violation of the weak energy condition arises in inflationary
cosmology.  In section~III we discuss whether a suitable integral
convergence condition might hold, even if the pointwise condition
does not.  Several integral conditions are known to give
rise to the focusing effects necessary for results such as
our previous theorems~[\BVOne\,\BVThree] to go through
\StoreRef{\Tipler}
\Refl{F.J. Tipler, \Jou{Phys. Rev.~D}, {\bf 17}, 2521 (1978).} 
\Refn{G. Galloway, \Jou{Manuscripta Math.}, \Vol{35}, 209 (1981).}
\Refn{T. Roman, \Jou{Phys. Rev.~D}, \Vol{33}, 3526 (1986);
     \Jou{Phys. Rev.~D}, \Vol{37}, 546 (1988).}
\StoreRef{\BordeFocus}     
\Refe{A. Borde, \Jou{Cl. and Quant. Gravity\/} \Vol{4}, 343 (1987).}. 
We argue, however, that even the weakest of the known integral 
conditions~[\BordeFocus] may not hold here.  
In section~IV we discuss the implications of the violation
of the weak energy condition for the existence of non-singular, eternally
inflating
cosmological models. We construct an explicit class of non-singular 
cosmologies, and we discuss why they are unsatisfactory as models
of eternal inflation.
We also discuss a property that realistic inflationary
scenarios might possess that would make all non-singular models
unsuitable as models of eternal inflation. 
In section~V we take stock of the situation: we compare
our approach to quantum stress-energy tensors with that of some
other authors, and we discuss the models
to which our earlier theorems~[\BVOne\,\BVThree] might still apply.

\Section Violation of the Weak Energy Condition

We first look at a simple model in which the inflating Universe is locally 
approximated by a Robertson-Walker metric:
$$
ds^2=a^2(\eta)(d\eta^2-d\vec x^{\,2}).\NumbEq
$$
\NameEq{\RW}
The approximation is justified when the scale of the spatial variation 
of the inflaton field~$\phi$ and of the Hubble parameter~$H$ is much greater 
than~$H^{-1}$.  The Hubble parameter is defined by
$H\equiv a'/a^2$ (where a prime is a derivative with respect to $\eta$)
and it obeys
$$
H'(\eta) = a^{-3} (aa'' - 2 a'^2).\NumbEq
$$
Consider a null vector of the form
$$
N^\mu = a^{-2}(1, \vec n), \quad \left|\vec n\right|=1, \NumbEq
$$
\NameEq{\Null}
where the ``normalizing factor'' $a^{-2}$ is chosen so as to ensure that
$N^\mu$ is the tangent to an affinely-parametrized
geodesic (a feature that we will need later). For such a vector, we have
$$
R_{\mu\nu}N^\mu N^\nu=-{2\over a^6}\Bigl(aa''- 2a'^2\Bigr)
=-{2\over a^3}H'=-{2\over a^2}\dot H.\NumbEq
$$
\NameEq{\WECExp}
A dot here is a derivative with respect to the proper time~$t$, related to
$\eta$ by $dt=ad\eta$.
Thus, in a region where $H' > 0$, the null convergence
condition will be violated.  

The Hubble parameter $H$ satisfies
$$
H^2 =  {8\pi G\over 3}\left({\>\dot\phi^2\over 2} + V(\phi)\right)
+ O(GH^4),\NumbEq
$$
where the last term represents the effect of the sub-horizon scale quantum
fluctuations we alluded to earlier (see eqn.~\Est). 
During inflation,
we have $\dot\phi^2\ll V(\phi)$, and if the energy scale of inflation is
well below the Planck scale, we also have $GH^2\ll 1$. The magnitude
of $H$ is then determined mainly by the inflaton potential $V(\phi)$.
In regions of deterministic slow roll, 
$$\bigl|\dot\phi\bigr| \approx
\left|V'(\phi)\right|/3H \gg H^2,\NumbEq$$ 
\NameEq{\DetSloRoll}
and
quantum fluctuations play a subdominant role in the dynamics of $\phi$.
In such regions, Einstein's equations with energy-momentum tensor for 
the averaged field are satisfied with good accuracy, and it is easily
verified that $\dot H \approx - 8\pi G\dot\phi^2 < 0$.
It then follows from~\WECExp\ that the weak energy condition is always 
satisfied in
slow-roll regions. On the other hand, in regions where the dynamics is 
dominated by quantum diffusion of the field $\phi$, eqn.~\DetSloRoll\
does not hold, and we have
$$
H^2 =  {8\pi G\over 3}V(\phi)+ O(GH^4).
$$
Quantum fluctuations take $\phi$ up and down the potential
$V(\phi)$, and the range of variation of $V(\phi)$ in the diffusion region 
is typically much greater than $H^4$. Hence, in some
parts of the diffusion region, $H$ will grow
and in other parts it will decrease.  The weak energy condition is
thus necessarily violated.

To see how this conclusion is affected by inhomogeneities of the
spacetime geometry, we consider a more general ansatz for the metric
$$
ds^2= a^2(\vec x, \eta) \bigl(d\eta^2 -  d\vec x^{\,2}\bigr).\NumbEq
$$
\NameEq{\GenRW}
For an inflating Universe with a slowly varying expansion rate
$H(\vec x, \eta)$, the scale factor has the form,
$$
a(\vec x,\eta)=\left(1 - \int_{\eta_0(\vec x)}^\eta H(\vec x,\hat\eta) 
d\hat\eta\right)^{-1}.
$$
With $N^\mu$ given by~\Null, we have 
$$
\eqalign{
R_{\mu\nu}N^\mu N^\nu&= 2 a^{-3} n^\mu n^\nu \partial_\mu \partial_\nu
\left({1\over a}\right)\cr
&=-2a^{-3}n^\mu n^\nu \partial_\mu \partial_\nu
\int_{\eta_0(\vec x)}^\eta H(\vec x,\hat\eta)\cr}\NumbEq
$$
\NameEq{\IntegrandTwo}
where $n^\mu = (1, \vec n)$.
To analyze the sign of this expression, we note that the scale factor
$a(\vec x, \eta)$ (and its inverse $a^{-1}$) may be 
expected to have many local
minima, maxima and saddle points as a function of $\vec x$ at any ``moment''
$\eta$. At such points, $\vec\grad a =0$ and \PrevEq\ can be written as
$$
R_{\mu\nu}N^\mu N^\nu = 2 a^{-3}\left[-{\partial H\over \partial \eta}
+ (\vec n\cdot{\vec\grad})^2\left({1\over a}\right)\right].
$$
At mimima of $a$, the second term in the square brackets is negative, and at
saddle points it is negative at least for some directions of $\vec n$. The first
term is negative whenever $H$ is increasing with time. In the diffusion region, 
we do not expect any strong correlations between the spatial dependence of the
scale factor (which is determined by the whole prior history of 
$H(\vec x, \eta)$) and the sign of $\partial H/\partial\eta$
(which depends only on the local quantum fluctuation of $H$).
Thus, it appears very likely that in some regions
both terms on the right-hand side will be negative and the weak energy condition
will be violated.

\Section Integral Convergence Conditions

The violation of the weak energy condition discussed above is not 
total: there are regions where the condition is violated, but also regions where it
is satisfied.  Moreover,
the probability for the field $\phi$ to move down the potential $V(\phi)$
is always greater than for it to move upward, and the weak energy condition is satisfied
when the field rolls down.  
This suggests that,
although there will be regions where the null convergence condition
will be locally violated, it may perhaps be satisfied in some
averaged sense.  

One kind of ``averaged'' condition is an integral 
convergence condition~[\Tipler--\BordeFocus].   
If we assume that an inflating spacetime is null-complete to the past,
then a past-directed null geodesic may be expected to cross regions
where the weak energy condition is satisfied as well as ones where it
may be violated.
Thus, it seems reasonable to ask whether an integral null convergence condition
along the lines
$$
\int R_{\mu\nu}N^\mu N^\nu dp \geq 0 \NumbEq
$$
\NameEq{\IntNullConv}
might hold, where the integral is taken along the geodesic, and
$p$ is an affine parameter with respect to which the tangent to the
geodesic, $N^\mu$, is defined (i.e., $N^\mu \equiv dx^\mu/dp$).
Condition \IntNullConv\ is either required to hold when the integral
is taken over the complete, or in some applications half-complete, geodesic
(as in the original proposal for such integral conditions~[\Tipler]),
or is required to ``repeatedly hold,'' as will happen when the
integrand oscillates~[\BordeFocus].

On examination, however, it is not clear if \IntNullConv\ will
hold when interpreted in either way. Consider, for example,
the metric~\RW.
Affinely parametrized null geodesics for this metric may be obtained
from the Lagrangian ${\cal L}=g_{\mu\nu}N^\mu N^\nu$.
The Euler-Lagrange equations reduce to
$d \bigl(a^2N^\mu\bigr) / dp=0$, 
where $p$ is an affine parameter, and we use the fact that 
$N^\mu$ is null.  One solution of this is the null vector
in~\Null.
For this solution, we have 
$$
dp=a^2d\eta=adt,\NumbEq
$$  
\NameEq{\AffPar}
and using~\WECExp\ we have
$$
\int R_{\mu\nu}N^\mu N^\nu dp = -2\int {H' \over a} d\eta 
= -2\int {\dot H \over a} dt. \NumbEq 
$$
\NameEq{\EvalNullInt}
The presence of~$a$ in the denominator makes the behavior of the
integral on the right difficult to assess. Without it~\PrevEq\ would reduce
merely to the difference in the values of H at the endpoints of
integration, and one could try and arrange for this difference to be positive along
at least some geodesics.  The presence of $a$ means, however, that there will
be increasingly larger contributions to the integral as we go to
earlier times (assuming that Universe is expanding) and it is not easy to
decide if the contributions of the wrong sign will always be compensated for
by those of the right sign.  
The situation is even more difficult in the case
of the more general metric~\GenRW. Here one would have to deal with the
integral of the complicated expression on the right-hand side of
\IntegrandTwo, and it is hard to see that one can argue that this
integral will either converge to a non-negative value, or even that
it will be ``repeatedly non-negative.''

\Section Non-singular Cosmologies

What are the consequences if, in addition to the pointwise violation of the
weak energy condition that we have discussed here, a suitable integral
condition also fails to hold?  One important consequence is that earlier
arguments that suggested that the Universe had a 
``beginning''~[\BVOne\,\BVThree] may no longer hold.
A crucial ingredient of these arguments is that a congruence of
initially converging geodesics comes to a focus. 
Convergence conditions, either pointwise or suitable integral ones,
guarantee focusing. Without such conditions, models can be constructed
where focusing does not occur, and in which geodesics can be extended
to infinite affine lengths in the past direction.

If a model based on the metric~\RW\ is to be non-singular, it follows from
\AffPar\ that
$$
\int_{-\infty}^t a(\tilde t) d\tilde t \NumbEq
$$
must diverge for all $t$, where $t$ is the proper time used above (defined via
$dt=ad\eta$).  We must also have $\dot a>0$ 
(for the Universe to be expanding).  
Cosmologies with a scale factor of the form $a(t)\sim (-t)^{-q}$, where
$0<q\leq 1$ (and $t<0$), satisfy these conditions. Such a scale factor 
appears, for example,
in the ``pre-big-bang'' stage of the
proposed models of string cosmology
\Refc{G. Veneziano, in \Jou{String Gravity and Physics at the 
Planck Energy Scale}, edited by N.\ Sanchez and A.\ Zichichi,
Kluwer Academic Publishers (1996).}
\Refe{M. Gasperini, in \Jou{String Gravity and Physics at the 
Planck Energy Scale}, edited by N.\ Sanchez and A.\ Zichichi,
Kluwer Academic Publishers (1996).}.
These models do not, however, qualify as models of ``steady state'' 
inflation.  The Riemann tensor in such models decreases as 
$R^{\mu\nu}_{\phantom{\mu\nu}\sigma\tau} \propto t^{-2}$ 
when $t\to -\infty$, indicating 
that the spacetime is asymptotically flat in the past direction.  
The Hubble parameter~$H$ also vanishes as $t\to -\infty$. This behavior
is very different from the quasi-exponential expansion with $H\approx 
\hbox{constant}$ that is characteristic of inflation at
later times.   Since the idea behind a steady-state model,
and its chief attraction, is that the Universe is in more-or-less 
the same state at all times, models with very different behaviour at 
early and late times are not viable as models of steady-state inflation.

Another example of a geodesically complete cosmology is
de~Sitter spacetime,
$$
ds^2= dt^2 - a^2(t) d\Omega_3^2, \NumbEq
$$
\NameEq{\deSitt}
where
$$
a(t)=H_0^{-1}\cosh(H_0t). \NumbEq
$$
For $t\gg H_0^{-1}$, the expansion rate $H$ is approximately equal
to the constant value $H_0$, and we have a canonical model of inflation.
This model, however, describes a contracting Universe for $t<0$.
Thermalized regions in such a Universe would rapidly merge and fill 
the entire space~[\Vilenkin].
The Universe would then collapse to a singularity and would never make it 
to the expanding stage. A further problem with a contracting Universe
is that it is extremely unstable. The growth of perturbations
by gravitational instability is slower in an expanding Universe than
in a flat spacetime, but in a contracting Universe the growth of 
perturbations accelerates.  Hence, a contracting Universe will rapidly 
reach a grossly inhomogeneous state from which it is not likely to
recover.

An inflating spacetime is not, of course, exactly de~Sitter, but is expected
to be locally close to de~Sitter. That is, for any spacetime point~$P$
there is a neighborhood of proper extent $\sim H^{-1}$ where the metric can be
brought to de~Sitter form with only small deviations from the exact
de~Sitter metric.  It has been
argued by one of us~[\Vilenkin] that such a spacetime is necessarily
contracting in the past, implying that steady-state inflation is impossible
in such a model.  That argument involved assumptions on the form of the
Riemann tensor. 
We provide here a new version of the argument based on the Ricci tensor.

Consider a congruence of timelike geodesics, past-directed from some
point~$p$.  Let the proper time, $t$, along these geodesics be zero at~$p$,
let $V^\mu$ be the tangent to the geodesics with respect to~$t$,
and let $\theta\equiv D_\mu V^\mu$ be the divergence of the congruence. If the
congruence is shear-free, as is the case in de~Sitter space, or in
general 2-dimensional spacetimes, we have
$$
{d\theta \over dt} = -{1 \over n-1} \theta^2 - R_{\mu\nu}V^\mu V^\nu,\NumbEq
$$
\NameEq{\RayN}
where $n$ is the spacetime dimension.  (This equation is a trivial extension of
the standard 4-dimensional geodesic focusing equation
\StoreRef{\HE}
\Ref{S.W. Hawking and G.F.R. Ellis, \Jou{The large scale structure of 
     spacetime}, Cambridge University Press, Cambridge, England 
     (1973).}.) 
Assume now that the Ricci tensor obeys
$$
R_{\mu\nu}V^\mu V^\nu < -{\delta^2 \over n-1} < 0\NumbEq
$$
\NameEq{\SECViol}
for all unit timelike vectors~$V^\mu$.
In other words, assume that the strong energy condition
is everywhere violated by at least a minimum amount. 
The strong energy condition requires that $R_{\mu\nu}V^\mu V^\nu \geq 0$ for 
all timelike vectors $V^\mu$, and this condition is violated in all
models of inflation that have been considered.
In fact, we have argued elsewhere~[\BVTwo] 
that a violation of 
this condition is necessary if a spacetime is to be considered ``inflating.''

Combining equations~\RayN\ and~\SECViol\ we get
$$
{d\theta \over dt} > {1 \over n-1} (\delta^2 - \theta^2).\NumbEq
$$
Integrating this expression from some negative value of $t$ (i.e.,
a point to the past of $p$ on the congruence) to zero, and 
using the fact that $\theta\to -\infty$ as $t\to 0^-$, we get
$$
\theta < \left[\coth \left({ t \over n-1}\delta\right)\right]\delta.\NumbEq
$$
Since $\coth x < -1$ for $x<0$,
eqn.~\PrevEq\ means that the past-directed timelike geodesics from $p$ 
continue to
diverge into the infinite past by an amount $\theta < -\delta$. Compare this
with flat space, where the geodesics also continue to diverge, but where
$\theta \sim 1/t$.  This faster-than-flat-space divergence suggests that the
Universe is contracting at early times. The objections raised above
then apply here as well.

Although suggestive, this argument 
falls short of a proof.  Congruences of geodesics
in realistic spacetimes will not
remain shear-free, and the effect of shear needs to be taken into account.
The global structure of locally de~Sitter
spacetimes remains an interesting open problem.

\Section Discussion

We have shown here that the weak energy condition generically will
be violated in inflating spacetimes.
Violations of the weak energy condition have been discussed by 
several other authors (see, for example, Flanagan and Wald
\StoreRef{\FW}
\Ref{\'E. Flanagan and R.M. Wald, \Jou{Physical Review~D}, \Vol{54},
6233 (1996).}
and references cited therein).
Previous work on the question has focused on the expectation value of the
energy-momentum tensor, $\langle T_{\mu\nu} \rangle$, and this approach 
has yielded limits on the violation of the weak energy condition.
In particular, Ford and Roman
\Refc{L.H. Ford, \Jou{Proc. R. Soc. London}, \Vol{A364}, 227 (1978);
\Jou{Phys. Rev. D}, \Vol{43}, 3972 (1991).}
\Refe{L.H. Ford and T.A. Roman, \Jou{Phys. Rev. D}, \Vol{41}, 3662 (1990);
     \Jou{Phys. Rev. D}, \Vol{44}, 1328 (1992);
     \Jou{Phys. Rev. D}, \Vol{51}, 4277 (1995);
     \Jou{Phys. Rev. D}, \Vol{53}, 1988 (1996);
     \Jou{Phys. Rev. D}, \Vol{53}, 5496 (1996).}
have investigated quantum states of free scalar and electromagnetic
fields in a flat spacetime for which $\langle T_{00} \rangle < 0$
in some region of spacetime.  They have shown that although such states
can be constructed, the magnitude of the negative energy density
and the time interval during which it occurs are limited by inequalities
that have the appearance of Uncertainty Principle inequalities.
Ford and Pfenning have obtained extensions of these ``quantum
inequalities'' to some curved spacetimes
\Refc{M. Pfenning and L.H. Ford, Tufts Institute of Cosmology preprint,
gr-qc/9608005 (1996).}
\Refe{M. Pfenning and L.H. Ford, to appear.}.
\StoreRef{\PF}
Flanagan and Wald~[\FW] have shown that an integral form of the weak
energy condition is satisfied for an appropriately smeared
$\langle T_{\mu\nu} \rangle$ in the case of a free, massless
scalar field in a nearly-flat spacetime.

Unfortunately, these results cannot be used to restrict the violations
of the weak energy condition of the type discussed in this paper.
One obvious reason is that the theorems proved so far are restricted to
free fields and a special class of spacetimes, which usually does not include
locally de~Sitter spaces.  (Pfenning and Ford have recently obtained
restrictions on violations of the weak energy condition in de~Sitter 
spacetime~[\PF],
but their results apply to a limited class of worldlines.)  
A more basic reason, however, is that
all these results are concerned with the expectation value 
$\langle T_{\mu\nu} \rangle$, while we are interested in the
fluctuations of $T_{\mu\nu}$.  The expectation value
$\langle T_{\mu\nu}(x) \rangle$ can be thought of as a result of
averaging the observed value of $T_{\mu\nu}$ at a point~$x$ in
an ensemble of identical Universes. In some of these Universes
the inflaton field will fluctuate ``up the hill,'' and the weak energy
condition will be violated, while in others it will go ``down the hill,''
and the condition will be satisfied.  Since the probability to go down is
always greater than probability to go up, we expect that on average
the weak energy condition will be satisfied; i.e., we expect that
$$
\langle T_{\mu\nu}(x) \rangle N^\mu N^\nu \geq 0. \NumbEq
$$
The violation of the weak energy condition, as well as the eternal 
character of inflation, both disappear when the field~$\phi$ and its 
energy-momentum tensor are replaced by their expectation values, since
both effects are due to relatively rare quantum fluctuations
of the field~$\phi$.

It is important to know if we can reasonably expect the energy conditions,
or suitable integral versions, to be satisfied,
because that will determine whether the singularity theorems, and other
results of classical general relativity,
will continue to hold (see Hawking and Ellis~[\HE]
for a review of these classical results and further references).
If previous singularity theorems that were aimed at
inflationary cosmology~[\BVOne\,\Borde\,\BVThree]
do not apply to some models,
we may then have the possibility of constructing
a ``steady-state'' eternally inflating Universe, without a beginning
and without an end.  The issue of whether or not the Universe
is past-eternal has been
discussed several times in the literature by Linde and his
collaborators
\Ref{See, for example, refs.~[\LindeOne], [\LindeTwo] and [\LindeThree].}.
They have argued that even when 
individual geodesics are past-incomplete, it may still be possible to view
the Universe as infinitely old
\Ref{A short discussion of this question also appears in
ref.~[\BVThree].}.  
What we have done here is
to point to a possibility~-- although a faint one~-- of constructing
fully non-singular models.  Such models, if they indeed exist, would
be geodesically past-complete.  

It must be noted that there is a class of inflationary models to which
our previous theorems do continue to apply.  These are models 
of ``open-universe'' inflation 
\Refl{J.R. Gott, \Jou{Nature}, \Vol{295}, 304 (1982).}
\Refn{M. Bucher, A.S. Goldhaber and N. Turok, \Jou{Phys. Rev. D},
      \Vol{52}, 3314 (1995).}
\Refn{K. Yamamoto, M. Sasaki and T. Tanaka, \Jou{Ap. J.}, \Vol{455}, 412 (1995).}
\Refe{A.D. Linde, \Jou{Phys. Lett.}, \Vol{B351}, 99 (1995).}
where the Universe consists of
post-inflationary ``bubbles'' embedded  
in a metastable false vacuum state.  Quantum diffusion of the inflaton
field does not occur here, 
and in the false vacuum the Ricci tensor is proportional to the metric.
In this case the null convergence condition is satisfied pointwise, 
and the models must possess initial singularities.

In other models of inflation,
we have shown here that there is a possibility
for non-singular models to exist, based on the violation of the weak energy condition
that occurs in these models. Whether realistic models of this type
can be constructed, however, remains open.
The discussion of Section~IV suggests that the construction of such models
may be difficult, if not impossible.

\Sectionvar Acknowledgements

The authors thank Larry Ford, Andrei Linde and Slava Mukhanov for discussions.
One of the authors (A.V.) acknowledges partial support from the National
Science Foundation. The other author (A.B.) thanks the Institute of
Cosmology at Tufts University and the High Energy Theory Group at
Brookhaven National Laboratory for their continued hospitality and support,
and the Research Awards Committee of Southampton College for partial 
financial support.

\EndPaper